# A Flower-Inspired Solution for Computer Memory Wear-Leveling

*Elizabeth Shen\*, Huiyang Zhou^, Senior Member, IEEE*
*\*Green Hope High School*
*^North Carolina State University*

*Abstract*— **Lengthening a computer memory's lifespan is important for e-waste and sustainability. Uneven wear of memory is a major barrier. The problem is becoming even more urgent as emerging memory (e.g., PCM) is subject to even shorter lifespan. Various solutions have been proposed, but they either require complicated hardware extensions or apply only to certain program constructs (e.g., loops).**

**This research proposes a new method, dual-ring wear leveling. It takes inspiration from the natural law known as the ``golden ratio" and how it helps flower petals evenly receive sun lights. By modeling memory as two rings and combines the idea with existing memory management, garbage collection (GC), the new solution offers an effective way to reduce memory wear and hence lengthen memory lifespan. It is deterministic, able to automatically adapt to memory size, requiring no hardware changes, and adding no slowdown to program executions.**

## I. INTRODUCTION

Computer memory is where data are stored during and after computations. It is subject to wearing. Each memory cell has a limited endurance. A cell of the latest phase-changing memory (PCM), for instance, is worn out after being written for about 109 times. Uneven wear among the cells in a memory is common due to the unequal chances for a memory cell to be accessed. It forces early retirement of a memory when many cells are still in good working condition. The issue causes the waste of billions of dollars each year for large data centers, and also leeches away at the money of the consumers.

Several potential solutions have been proposed [8,9,10,11,12,13]. A good portion of these proposals [8,12] involve aging awareness, which is when the current cell age, or number of times it's been written, is taken into account when choosing which cell to write to.

However, modern day memory hardware does not include a count of how frequently a cell has been written, which warrants additional hardware to be required. It would increase the complexity and cost, as well as runtime overhead; how practical it is yet to be determined. Furthermore, this technique would prove to be much slower and less efficient than current computer memory, and would not be able to withstand the high intensity of demands placed upon computer memory today.

Another suggested solution is to alter any recursive loops to prevent the repeated access of certain individual cells [13]. While effective in preventing wear caused by recurring loops, this solution does nothing to address the numerous other ways memory wear occurs, hence failing to level memory wear in a substantial way.

This paper proposes a new way to address the uneven wear problem of memory. Flower petals came as the source of inspiration for our solution.

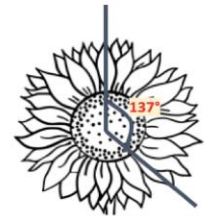

**Fig 1.** Golden ratio seen in flower petals.

In nature, a phenomenon known as the golden ratio, can be found being used to determine the placement or structure of countless items. Most notably are flower petals that can be viewed to be placed every 137 degrees rotation from the last one, as shown

Figure 1. The golden ratio is also referred to as Fibonacci's code, and is defined by 137 degrees of rotation in Geometry and a fixed ratio (about 1.618) between parts. The rule produces a spiral that perfectly distributes each component. In the case of flower petals, the petals are placed in a way that ensures there is little overlap, and the sun is able to access all of them equally.

The key finding in this work is that the golden ratio retains its benefits when applied to a computer science environment, helping level the accesses on computer memory.

Not only is this solution effective, it's also deterministic, a factor that will benefit reproduction, debugging, and diagnosis of issues. There are no limits to the size of memory this solution can be applied to, and is wholly adaptable to better suit various memory including future ones if the need arises.

The actual process of realizing the idea however proved to be challenging. Two important questions need to be solved: (i) how to translate the principal of the golden ratio into a mechanism that effectively levels memory accesses; (ii) how to make the mechanism work.

The way the golden ratio evens exposure to sunlight on flower petals relies on the subject to be circular, so a rotation of 137 degrees would be possible. Computer memory is however inherently linear. To address this first challenge, we created a dual-ring, generational scheme, by treating the full memory as a collection of two halves, with each half logically modeled as a ring. The two rings are used as working memory alternatively and periodically. When the period comes for a ring to be the working memory, all the useful data on the other ring is copied to it. The start location where the data is written on the ring is a 137-degree rotation away from the location used at the beginning of the previous working period of this ring. By shifting the location of data in a way similar to how flower petals are placed, the accesses and hence wear of the memory cells are evened out.

To address the second challenge, the runtime overhead problem, we integrated the solution into existing memory management. Copying data back and forth adds substantial overhead to the execution time of a program. Our solution is to embed the mechanism into the existing memory management schemes in computer programs, software garbage collection (GC). GC plays an important role in modern programs in periodically reclaiming the memory space occupied by useless data. In that process, it usually packs together the useful data on memory such that free space can be in big chunks, easier to use for new data. Our solution integrates the dual-ring generational wear leveling mechanism into the data packing step in GC, which not only evens access to memory but also avoids extra data copying operations.

We evaluated the solution on both synthetic and real-world memory access traces. Our experiments on the traces of seven real-world programs show that the dual-ring generational wear leveling method is effective in making the average access each cell receives 1.8 to 6.3 times less, and the max accesses 1.1 to 13.2 times less. It adds no extra time to the default program executions. As a pure software solution, this new method requires no hardware changes. Its inheritance of the appealing properties of golden ratio makes it automatically adapt to the memory size, data object size and access frequency. In addition, it is also deterministic, allowing easy reproduction and diagnosis of program executions.

## II. BACKGROUND

**Golden Ratio**  It is also known as the golden section, golden mean, or divine proportion. In mathematics, it is calculated as the irrational $(1 + \sqrt{5})/2$, often denoted by the Greek letter ϕ or τ, which is approximately equal to 1.618. In Britannica, it is defined as "the ratio of a line segment cut into two pieces of different lengths such that the ratio of the whole segment to that of the longer segment is equal to the ratio of the longer segment to the shorter segment". Geometrically, it corresponds to an about 137-degree rotation in a circle. The origin of this number can be traced back to Euclid, who mentions it as the "extreme and mean ratio". In nature, one of the most fundamental (and strikingly beautiful) ways mathematics laws manifest is through the golden ratio. When the golden ratio is applied as a growth factor, you get a type of logarithmic spiral known as a golden spiral. Examples in nature are flower petals, pinecones, seashells, Romanesco broccoli, whirlpools, even galaxy.

**Garbage Collection (GC)**  GC is a common mechanism for managing memory in a computer during the execution of a program in modern programming languages. During the execution of a program, memory is continuously allocated to hold some useful data for computations and freed when they are done being used. GC is automatically invoked by the computer system. It scans through the memory, marks useful data on memory, packs the data up to a continuous chunk of memory, and reclaims the freed memory space. Without GC, the memory may quickly run out and there could be many small free spaces scattered on the memory, making it hard to hold a large data object.

## III. RELATED WORK

Previous years have brought numerous studies and research on the issue of memory wearing, and its possible solutions. Aside from the aging-aware technique, which takes into account the amount a cell has been written, there are also the random-based approaches.

These include the application of wear-leveling in a circular or random format [8,12]. Others proposed variations to both techniques, such as a software-only aging-aware wear-leveling, in which the aging-aware approach was implemented into the Memory Management Unit (MMU), and an internal access counter was embedded into the system. In this version, the worn-down cells, or the cells with the most accesses, are swapped out for cells with the least accesses each time a GC is called [9,11]. However, while this solution does aid in wear-leveling, it increases overhead time with a 7.2% longer mutator execution time.

Another suggested approach was a solution where a compiler changes loops with recursive functions to prevent loops from calling the same memory area repeatedly [13]. While this does solve one source of the memory wear, it fails to address the numerous other causes, and does little to prevent the rapid wearing of the memory. This solution is also inflexible, and is difficult to scale and implement across a variety of compilers. The program is also unable to be generalized to suit all programming languages, including those of the future.

## IV. DESIGN

This section presents our solution, namely petal memory. It is a pure software solution, requiring no hardware changes. Its core technique is a dual-ring wear leveling mechanism. We next first explain our dual-ring model of computer memory, and then describe the wear leveling mechanism, along with an example.

*A. Dual-Ring Memory Model*

The source of inspiration, flower petals, enjoy the even exposure to sunlight by growing the petals according to the golden ratio; the next petal to grow is 137-degree away from the previous petal. To leverage this mechanism for memory, the first problem is that the computer memory is a linear space, going from the first cell with address 0 to the last cell with the largest address, as illustrated in the top part of Figure 2.

In our design, we created a logical view of the memory space as two rings. In this model, the memory is viewed as two equal-sized chunks, as shown in the middle part of Figure 2. Then, for each of the two chunks, we view it as a ring, with the head of the chunk as the next cell of the end of the chunk on the ring, illustrated in the bottom part of Figure 2.

Modeling the linear memory space into two rings puts each half of the linear memory space into a form ready for the application of petal-inspired data placement on memory.

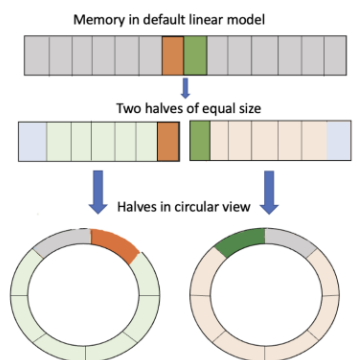

**Fig 2**. Dual-ring modeling of the linear computer memory.

*B. Dual-Ring Wear-Leveling Algorithm*

Based on the dual ring model of memory, we designed the following algorithm to apply golden ratio to level memory wear. The algorithm is integrated into existing GC such that the leveling consumes no extra time. Figure 3 shows the algorithm with a pseudo code. It uses "Mem1" and "Mem2" to represent the two rings of memory. At the beginning, it lets "Mem1" be the work memory, "Mem2" be the idle memory (Lines 1-3 in Figure 3). All following memory operations (allocation, deallocation, reads, writes) all happen on the work memory. As the background section has mentioned, GC is invoked periodically during the execution of a program. The dual-ring wear leveling takes place during a GC. During a GC, the useful data are packed together and moved from the work memory to the idle memory (Lines 9 to 12 in Figure 3). The start destination location on the idle memory is determined based on the golden ratio by calling the function nextPetal on Line 7. The implementation of nextPetal is shown in Lines 15 to 21 in Figure 3. The algorithm splits the idle memory into two parts, the longer, A, and the shorter, B. We applied the golden ratio equation as follows:

$$\frac{A+B}{A} = \frac{A}{B} = 1.618$$

where, A+B is the total length of the idle memory ring. So A can be computed as the length of the idle memory divided by 1.618, and B is the difference of A from the total length of the idle memory ring. B is how far the next petal should be from the previous petal in a flower so that the rotation is 137 degrees. It is how nextPetal determines the new location for the GC to use to place the useful data on the idle memory (Line 18 in Figure 3). When the result may exceed the memory size, the algorithm automatically wraps around the ring (Lines 19, 20 in Figure 3).

At the end of a GC, the work memory is cleaned, and the roles of the work memory and the idle memory switches so that the next period of memory operations will happen on the new work memory (i.e., the previous idle memory).

As the GC shifts the memory locations where data are copied, the entire memory gets a more leveled chances of being accessed, just like how flower petals receive similar amount of sunlight.

```
Pseudo Code of Dual-Ring Wear Leveling
1.  # Initial setting
2.  workMem = Mem1
3.  idleMem = Mem2

4.  # At GC time
5.  GC( ):
6.     prevPetalLoc = copytoStartLoc[idelMem]
7.     petalLocs [idleMem] = nextPetal (prevPetalLoc)
8.     copytoLoc = petalLocs[idleMem]
9.     scan workMem:
10.       if (a useful data chunk C is found):
11.          copy C to copytoLoc on idleMem
12.          copytoLoc = copytoLoc + size(C)
13.    clean workMem
14.    swap workMem and idelMem

15. # Implementation of nextPetal
16. # φ is the golden ratio 1.618
17. nextPetal(prevPetal):
18.    newLoc = prevPetal + memorySize(1-1/φ)
19.    if (newLoc > memorySize):
20.       newLoc = newLoc - memorySize
21.    return newLoc
```

**Fig 3**. Pseudo code of the dual-ring wear leveling algorithm.

*C. Example*

To help with the understanding, Figure 4 illustrates how the dual-ring wear leveling method works. The left column shows the status changes of the first ring (Mem1), and the right column shows the status changes of the second ring (Mem2). At the first GC, the useful data are copied to the head section of the idle memory (top right of Mem2). Then the roles of the two memory switches and the program execution allocates some new data and deallocates some old data on Mem2 (right second row in Figure 4). When another GC happens, the useful data are copied to Mem1 where the start location (marked as petal location in Figure 4) is 137-degree rotation away from the previous petal location (left second row in Figure 4). This process continues. In both columns in Figure 4, we can see that the sections of the memory being actively operated on shifts based on the golden ratio, evenly spread out across the memory.

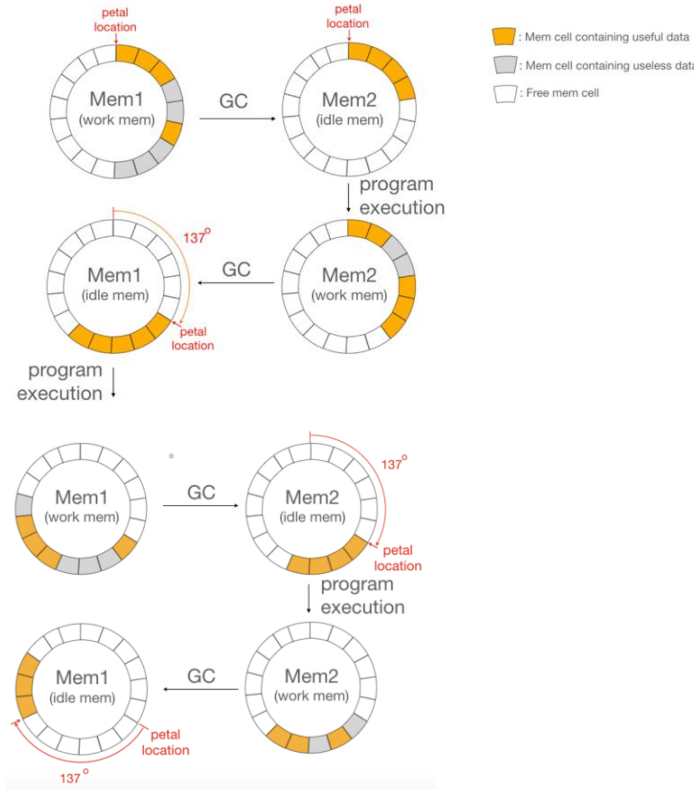

**Fig 4**. Illustration of dual-ring wear leveling in action

### D. Analysis

The use of golden ratio makes the solution adapt to the memory size, data size, and access patterns. For the nature of golden ratio, the placement of data on memory is evenly distributed regardless of those factors, just like how flower petals are evenly distributed regardless of how large the petals are. In comparison, if the shift is a constant, the shift may be too small for some memory and too large for another. Another appealing property of our solution is that it is deterministic. If we change the algorithm such that each time GC takes a random shift, the memory accesses may also be close to even, but the execution becomes non-deterministic, introducing difficulties for testing, diagnosis, and reproduction of the executions.

## V. EXPERIMENTAL RESULTS

This section reports the experiments We have done to evaluate the proposed technique. The questions we try to answer are as follows:

- How effective is the overall dual-ring wear leveling method in reducing the uneven wear of memory?
- How useful is it to use golden ratio in the solution?
- How much time overhead does the method introduces to program executions?

### A. Methodology

**Table 1**. Real-world programs for experiments and their memory trace lengths.

| Name | Description | Trace Length(entries) |
|---|---|---|
| bzip2 [1] | Data compressor. 2x faster at compressing. 6x faster at decompressing | 123.2 mil. |
| du [2] | A command that will display the amount of disk space used by a specific file or directory. | 24.4 mil. |
| espeak [3] | Text to speech software. Uses 'formant synthesis' methodology. | 37.9 mil. |
| nab [4] | Molecule manipulation | 38 mil. |
| python [5] | A computer programming software. | 12.1 mil. |
| x264 [7] | Video encoding | 34.4 mil. |
| xml2json [8] | Converts an XML message into JSON with the same structure. | 530k |

To answer these questions, we have implemented the dual-ring wear leveling algorithm in Python (v3.6.5). Our program takes a memory trace as input, enumerates the memory operations in the trace while applying the wear leveling algorithm, and reports the numbers of times each memory cell gets accessed. A memory trace is a sequence of memory operations recorded during a program execution, including memory allocation, deallocation, reads, writes, and GCs.

The memory traces of seven real-world programs were collected with Intel PIN tool[14] when the programs run on an Intel Xeon machine. These programs are selected from several domains to form a comprehensive set. As listed in Table 1, they include a data compression utility program, a disk space checking program, a text-to-speech program, a molecule manipulation program, the standard Python interpreter (v3.6), a video encoder, and an XML-to-JSON file converter. The lengths of the traces used in the experiments range from 530K entries to 123M entries as the right column in Table 1 shows. I, in addition, measured the GC time with and without dual-ring wear leveling; there is no statistically significant difference, indicating that the introduction of our solution adds no time overhead.

### B. Results

We compared our proposed dual-ring wear leveling (denoted as gold) with the default execution of the programs (denoted as default), and a variation of dual-ring wear leveling, which shifts a quarter of the memory (shiftQuarter) rather than the amount determined by golden ratio.

6991019

Table 2 reports the average and the maximum numbers of accesses to a memory cell when each of the programs run with each of the three memory wear leveling schemes. The two variants of the dual-ring wear leveling scheme, golden and shiftQuarter, can both reduce the numbers of accesses per memory cell significantly, while golden consistently performs the best, indicating the benefits of the use of golden ratio.

**Table 2.** Average and maximum number of accesses to a memory cell

| Program | Average # accesses per cell | | | Maximum # accesses to a cell | | |
|---|---|---|---|---|---|---|
| | golden | default | shiftquarter | golden | default | shiftquarter |
| bzip2 | 3423 | 5170 | 4696 | 9332 | 50970 | 13154 |
| du | 243 | 730 | 587 | 2054 | 9303 | 2706 |
| espeak | 426 | 1997 | 830 | 2158 | 12724 | 4204 |
| nab | 2812 | 5117 | 5059 | 3494 | 170234 | 4384 |
| python | 460 | 818 | 642 | 1483 | 23254 | 6180 |
| x264 | 110 | 694 | 249 | 741 | 2657 | 847 |
| xml2json | 32 | 69 | 47 | 98 | 1290 | 344 |

Figure 5 provides an in-depth view of the results on two of the programs, bzip2 and du. Each of the violin graphs shows the distribution of the number of accesses among the 1000 most frequently accessed memory cells under each of the three schemes. The golden scheme reduces the accesses to the most frequently accessed cell from 50,970 down to 9332, and the overall distribution becomes much narrower. The ShiftQuarter scheme also reduces the distribution, but not as much as the golden scheme does. The graphs on du show a similar result.

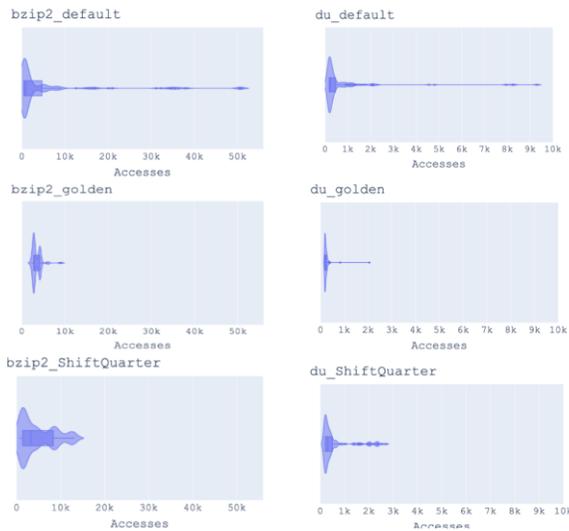

**Fig 5**. Distribution of the numbers of accesses of program bzip2 under three methods

Figure 6 shows how much the lifespan of the memory increases based on the reduction of the average number of accesses and the maximum number of accesses. The golden scheme extends the average lifespan by 1.5X--6.3X and the lifespan of the most frequently accessed cells by 3.6X--48X. The ShiftQuarter scheme also helps, but much less.

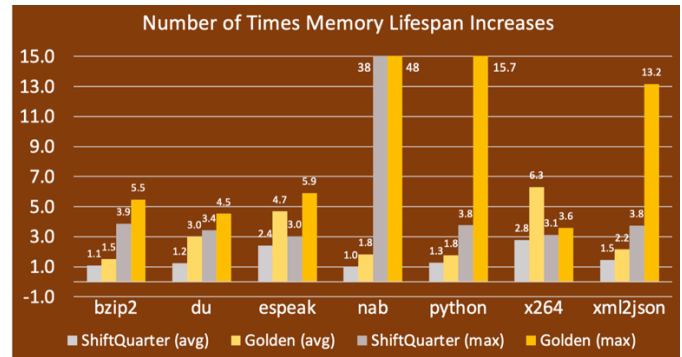

**Fig 6**. Extensions of the lifespan of memory calculated by the numbers of reduced accesses

## VI. CONCLUSION

This research shows that the proposed solution, dual-ring wear leveling, can significantly reduce memory wear. It also significantly outperforms the variation where golden ratio is not used, which confirms the value of golden ratio in the proposed solution. The solution is software-only, requiring no hardware changes. It is not limited to any types of program structure or memory size, showing significant benefits across all experimented programs. It is at the same time deterministic. Its integration in GC ensures little to no overheard processing time is added, and provides the same leveling benefits that the inspiration first promised.

A direction worth future exploration is to reduce the accesses to the most frequently visited memory cells. In a program, there are always some data used repeatedly, much more frequently than other data, such as those outliers in Figure 5. To reduce memory wear caused by accesses to those data, some other methods may need to be invented to complement the dual-ring wear leveling method. Besides memory, the technique is potentially applicable to system-level storage; which is worth studying in the future.

6991019

REFERENCES

[1] *bzip2*. (2019, January 21). Retrieved from sourceware: https://www.sourceware.org/bzip2/
[2] Reynolds, L. (2021, September 16). Retrieved from LinuxConfig: https://linuxconfig.org/du
[3] *eSpeak text to speech*. (2021, September 17). Retrieved from Sourceforge: http://espeak.sourceforge.net/
[4] Tom Macke, W. S.-S. (2021, December 19). *The NAB molecular manipulation language*. Retrieved from CaseGroupRutgers: https://casegroup.rutgers.edu/casegr-sh-2.2.html
[5] *About Python*. (n.d.). Retrieved from python: https://www.python.org/
[6] Project, V. (2021, December 31). x264 *Video Codec*. Retrieved from majorgeeks: https://www.majorgeeks.com/files/details/x264_video_codec.html
[7] Retrieved from ubuntu manuals: http://manpages.ubuntu.com/manpages/impish/man1/xml2json.1p.html
[8] Alexandre P. Ferreira, M. Z. (2010). *Increasing PCM main memory lifetime*. Leuven: European Design and Automation Association.
[9] Chengwen Wu, G. Z. (2016). *Rethinking Computer Architectures and Software Systems for Phase-Change Memory*. New York: Association for Computing Machinery.
[10] Christian Hakert, K.-H. C.-J. (2020). *Software-Only In-Memory Wear-Leveling for Non-Volatile Main Memory*. arXiv arXiv:2004.03244.
[11] Lingyu Zhu, Z. C. (2018). *Wear Leveling for Non-Volatile Memory: a Runtime System Approach*. IEEE.
[12] Ping Zhou, B. Z. (2009). *A durable and energy efficient main memory using phase change memory technology*. New York: Association for Computing Machinery.
[13] Wei Li, L. W. (2020). *Loop2Recursion: Compiler-Assisted Wear Leveling for Non-Volatile Memory*. IEEE.
[14] CK Luk, R. C. (2005). in Proceedings of the 2005 ACM SIGPLAN congerence on Programming language design and implementation.